\documentclass[aps,prc,preprint,showpacs,showkeys]{revtex4}

\usepackage{graphicx}
\usepackage{bm}
\usepackage{epsf}

\begin{document}

\title{Pentaquark $\Theta^+$ in nuclear matter and $\Theta^+$ hypernuclei}

\author{H. Shen}
\email{songtc@public.tpt.tj.cn}
\affiliation{Department of Physics, Nankai University, Tianjin 300071, China}

\author{H. Toki}
\email{toki@rcnp.osaka-u.ac.jp}
\affiliation{Research Center for Nuclear Physics (RCNP), Osaka University,
             Ibaraki, Osaka 567-0047, Japan}

\begin{abstract}
We study the properties of the $\Theta^+$ in nuclear matter and
$\Theta^+$ hypernuclei within the quark mean-field (QMF) model,
which has been successfully used for the description of ordinary
nuclei and $\Lambda$ hypernuclei. With the assumption that the
non-strange mesons couple only to the $u$ and $d$ quarks inside
baryons, a sizable attractive potential of the $\Theta^+$ in
nuclear matter is achieved as a consequence of the cancellation
between the attractive scalar potential and the repulsive vector
potential. We investigate the $\Theta^+$ single-particle energies
in light, medium, and heavy nuclei.  More bound states are
obtained in $\Theta^+$ hypernuclei in comparison with those in
$\Lambda$ hypernuclei.
\end{abstract}



\maketitle

\section{Introduction}
\label{sec1}

The signature of the pentaquark was found in the experiment with
GeV photons on $^{12}$C at LEPS of
SPring8~\cite{nakano02,nakano03}. The $\gamma n \rightarrow K^+
K^- n$ reaction was analyzed and a narrow resonance state was
identified with the mass of $1.54\;\rm{GeV}$ and the width smaller
than $25\;\rm{MeV}$. Such a narrow state was predicted by Diakonov
et al.~\cite{diakonov97} as a pentaquark state at nearly the same
energy.  There were many experimental works published after the
original study by using various reactions~\cite{penta04}.
Recently, many further experiments, particularly at very high energy,
report negative results except for a few positive results~\cite{penta04}.
At this moment, the spin and parity of the state are not known.
Even the width is not precisely identified experimentally.

There are many theoretical works~\cite{penta04}.  It seems those
models, which emphasize the importance of the chiral symmetry,
suggest the positive parity ($J^P=1/2^+$) as the original
predication of Diakonov et al.~\cite{diakonov97}. On the other
hand, the naive constituent quark models tend to provide the
negative parity ($J^P=1/2^-$).  The QCD originated works as
the lattice QCD and the QCD sum rule approaches seem to provide
the negative parity, although these methods face the difficulty
of handling the nearby $K^+n$ threshold and the small pion mass.
Hence, we are at this moment not sure about the presence of the
pentaquark state, the spin-parity and even the width of this
state~\cite{penta04}.

We should make further efforts to get more information for the
pentaquark state.  There are several interesting theoretical works
to study the pentaquark in nuclear medium~\cite{Miller,Kim,Oset1,Tsushima,Zhong}.
In Ref.~\cite{Oset1}, the authors study the interaction of the
pentaquark with nucleons using the meson exchange model. They have
obtained a large binding potential with the width to be small to
take the pentaquark nucleus seriously. This work then motivated a
theoretical work to calculate the ($K^+,\pi^+$) reaction for the
formation of the pentaquark nuclei~\cite{Oset2}.  There is a good
possibility to see $\Theta^+$ hypernucleus in such a reaction.
It is then very interesting to study the $\Theta^+$ hypernucleus
from different view point.

We studied nuclei and hypernuclei in terms of the quark mean-field
(QMF) model~\cite{qmf2,qmf3}.  We assume in the QMF model that up
and down quarks in the baryons interact with $\sigma$, $\omega$,
and $\rho$ mesons (mesons made of up and down quarks) in nuclear
matter and in nuclei. Hence, the interaction of the $\Theta^+$
with other nucleons is obtained without introducing new parameters
in the QMF model.  The only assumption needed here is the isospin
of the $\Theta^+$ to be $I=0$.  The interactions of quarks with
the meson fields change the properties of the baryons in hadronic
matter and hence the hadronic matter property is obtained from
these interactions self-consistently.

In the next section, we would like to present the formulation of
the $\Theta^+$ hypernuclei in the relativistic quark mean-field
model.  In section~\ref{sec3}, we provide the numerical results for
various $\Theta^+$ hypernuclei.  Section~\ref{sec4} will be
devoted to the conclusion and discussions.

\section{Quark mean-field model for the $\Theta^+$ in medium}
\label{sec2}

We start with the description of the $\Theta^+$ in nuclear medium,
which is comparable to the treatment of $\Lambda$ hypernuclei in
our previous work~\cite{qmf3}. The $\Theta^+$ is known to be an
exotic baryon containing five quarks ($uudd\bar{s}$) with
strangeness $S=+1$, isospin $I=0$, and charge $Q=+1$.  However,
its parity is still ambiguous experimentally. In this work, we use
the constituent quark model to describe the $\Theta^+$ and
nucleons in medium, while the exchanged mesons couple directly to
the quarks inside baryons. The constituent quarks satisfy the
Dirac equation:
\begin{eqnarray}
\label{eq:dirac}
\left[i\gamma_{\mu}\partial^{\mu}-m_q-\chi_c
-g^q_{\sigma}\sigma
-g^q_{\omega}\omega\gamma^0
-g^q_{\rho}\rho\tau_3\gamma^0
\right] \phi^q(r) = 0,
\end{eqnarray}
where $\chi_c$ is the confining potential taken as the quadratic
form, $\chi_c=\frac{1}{2}kr^2$. Because the Lorentz structure of
the confining interaction is not well known yet, we here use a
scalar confinement. In our previous studies for finite nuclei and
$\Lambda$ hypernuclei~\cite{qmf2,qmf3}, both scalar and
scalar-vector confinements have been taken into account, but it is
known that antiquarks cannot be confined by a scalar-vector
confinement when the vector part is equal or larger than the
scalar part. Therefore we just adopt the scalar confinement in the
present calculation of $\Theta^+$. We assume that the non-strange
mesons, $\sigma$, $\omega$, and $\rho$, couple exclusively to the
up and down quarks and not to the strange quark (or antiquark)
according to the OZI rule, hence the strange antiquark inside
$\Theta^+$ satisfies the Dirac equation in which the terms coupled
to the non-strange mesons vanish
($g^{s,\bar{s}}_{\sigma}=g^{s,\bar{s}}_{\omega}=g^{s,\bar{s}}_{\rho}=0$).
We note that the Dirac equation for the strange antiquark
$\bar{s}$ in $\Theta^+$ is identical to the equation for the
strange quark $s$ in $\Lambda$ when a scalar confining potential
is adopted in Eq. (\ref{eq:dirac}). We follow Ref.~\cite{qmf2} to
take into account the spin correlations and remove the spurious
center of mass motion, and then obtain the effective mass of the
$\Theta^+$ in medium as
\begin{eqnarray}
\label{eq:mstart}
M_{\Theta}^{*}=\sqrt{(4e_q+e_{\bar{s}}+E^{\Theta}_{\rm{spin}})^2-
                     (4\langle p_q^2 \rangle+\langle p_{\bar{s}}^2 \rangle)},
\end{eqnarray}
while the effective mass of the nucleon is expressed as
\begin{eqnarray}
\label{eq:mstarn}
M_N^{*}=\sqrt{(3e_q+E^N_{\rm{spin}})^2-
               3\langle p_q^2 \rangle }.
\end{eqnarray}
Here, the subscript $q$ denotes the $u$ or $d$ quark. The energies
($e_q$ and $e_{\bar{s}}$) and momenta ($\langle p_q^2 \rangle$ and
$\langle p_{\bar{s}}^2 \rangle$) can be obtained by solving the
corresponding equations. The spin correlations
($E^{\Theta}_{\rm{spin}}$ and $E^N_{\rm{spin}}$), which are taken
as parameters in the QMF model, are determined by fitting the
baryon masses in free space ($M_{\Theta}=1540\;\rm{MeV}$ and
$M_N=939\;\rm{MeV}$). The effective masses of baryons are
influenced by the $\sigma$ mean-field in nuclear medium, which
provides a scalar potential to the $u$ and $d$ quarks in baryons
and as a consequence reduces the constituent quark mass to
$m_q^{*}=m_q+g_{\sigma}^q\sigma$ ($q=u,d$). However, the $\omega$
and $\rho$ mean fields could not cause any change in the baryon
masses, and they appear merely as the energy shift. Therefore, we
obtain the effective masses of baryons as functions of the
$\sigma$ mean-field, $M_\Theta^{*} (\sigma)$ and $M_N^{*}
(\sigma)$, expressed by Eqs. (\ref{eq:mstart}) and
(\ref{eq:mstarn}).

We now derive a self-consistent treatment for a $\Theta^+$ hypernucleus,
which contains a pentaquark $\Theta^+$ and many nucleons.
The baryons in this system interact through the exchange of
$\sigma$, $\omega$, and $\rho$ mesons, just like in the case of a
single $\Lambda$ hypernucleus. The effective Lagrangian at the hadron
level within the mean-field approximation can be written as
\begin{eqnarray}
\label{eq:lag}
{\mathcal L} &=&
\bar\psi\left[ i\gamma_\mu\partial^\mu-M_N^*(\sigma)
-g_\omega \omega \gamma^0
-g_\rho \rho \tau_3\gamma^0
-e\frac{(1+\tau_3)}{2} A \gamma^0
\right] \psi  \\ \nonumber
 & &
+\bar\psi_{\Theta} \left[ i\gamma_\mu\partial^\mu-M_{\Theta}^*(\sigma)
-g^{\Theta}_\omega \omega \gamma^0
-e A \gamma^0
\right] \psi_{\Theta} \\ \nonumber
 & &
-\frac{1}{2} (\bigtriangledown\sigma)^2
-\frac{1}{2} m_\sigma^2\sigma^2
-\frac{1}{4} g_3\sigma^4
+\frac{1}{2} (\bigtriangledown\omega)^2
+\frac{1}{2} m_\omega^2\omega^2
+\frac{1}{4} c_3\omega^4   \\ \nonumber
 & &
+\frac{1}{2} (\bigtriangledown\rho)^2
+\frac{1}{2} m_\rho^2\rho^2
+\frac{1}{2}(\bigtriangledown A)^2,
\end{eqnarray}
where $\psi$ and $\psi_{\Theta}$ are the Dirac spinors for the
nucleon and $\Theta^+$. The mean-field values are denoted by
$\sigma$, $\omega$, and $\rho$, respectively, while $m_{\sigma}$,
$m_{\omega}$, and $m_{\rho}$ are the masses of these mesons.
The electromagnetic field, denoted by $A$, couples both to the
proton and $\Theta^+$, which carry a positive charge.
The influences of the $\sigma$ mean-field on baryons are
contained in the effective masses $M_N^{*} (\sigma)$ and
$M_\Theta^{*} (\sigma)$. The $\omega$ meson, which couples
directly to the $u$ and $d$ quarks inside baryons, provides
an interaction at the hadron level with the coupling constant
$g^{i}_\omega=n_i g_\omega^q$, where $n_i$ is the number of
the $u$ and $d$ quarks in the baryon $i$.
When $i=\Theta^+$ ($i=N$), we get $g^{\Theta}_\omega=4g_\omega^q$
($g^N_\omega=g_\omega=3g_\omega^q$). The $\rho$ meson doesn't
couple to the $\Theta^+$ which is an isoscalar particle,
but it couples to the nucleon with the coupling constant
$g^N_\rho=g_\rho=g_\rho^q$, as given in Ref.~\cite{qmf2}.
In the QMF model, the basic parameters are the quark-meson couplings
($g^q_\sigma$, $g_\omega^q$, and $g_\rho^q$), the nonlinear
self-coupling constants ($g_3$ and $c_3$), and the mass of the
$\sigma$ meson ($m_\sigma$), which have been determined by fitting
the properties of nuclear matter and finite nuclei in Ref.~\cite{qmf2}.
Therefore, no more adjustable parameters exist
when it is extended to the calculations for $\Theta^+$ or
$\Lambda$ hypernuclei. From the Lagrangian given in Eq. (\ref{eq:lag}),
we obtain the following Euler-Lagrange equations:
\begin{eqnarray}
 & & \left[
i\gamma_{\mu}\partial^{\mu}-M_N^*(\sigma)
-g_\omega \omega \gamma^0
-g_\rho \rho \tau_3\gamma^0\
-e\frac{(1+\tau_3)}{2} A \gamma^0
 \right]\psi
= 0, \\
 & & \left[
i\gamma_{\mu}\partial^{\mu}-M_\Theta^*(\sigma)
-g^\Theta_\omega \omega \gamma^0
-e A \gamma^0
 \right]\psi_\Theta
= 0,
\\
 & & \left(
 -\Delta+m_\sigma^2\right) \sigma=
-\frac {\partial M_N^*(\sigma)}{\partial \sigma} \rho_s
-\frac {\partial M_\Theta^*(\sigma)}{\partial \sigma} \rho^\Theta_s
-g_3 \sigma^3,
\\
 & & \left(
 -\Delta+m_\omega^2\right) \omega=
g_\omega \rho_v +g^\Theta_\omega \rho^\Theta_v
-c_3 \omega^3,
\\
 & & \left(
 -\Delta+m_\rho^2\right) \rho =
g_\rho \rho_3,
\\
 & &
 -\Delta A =
e \left( \rho_p+\rho^\Theta_v\right),
\end{eqnarray}
where $\rho_s$ ($\rho_s^\Theta$), $\rho_v$ ($\rho_v^\Theta$),
      $\rho_3$, and $\rho_p$
are the scalar, vector, third component of isovector, and proton
densities, respectively. The above coupled equations are solved
self-consistently with the effective masses $M_N^{*}(\sigma)$ and
$M_{\Theta}^{*}(\sigma)$ obtained at the quark level.

The application of the QMF model to the description of $\Lambda$
hypernuclei has been reported in our previous work~\cite{qmf3}.
Without adjusting any parameters, the properties of $\Lambda$
hypernuclei could be described reasonably well.
The $\Lambda$ single-particle energies obtained in the QMF model
are slightly underestimated in comparison with the experimental values.
In the present work, we investigate the $\Theta^+$ hypernuclei
within the QMF model, and then provide the predictions for
$\Theta^+$ single-particle energies in light, medium, and heavy nuclei.

\section{Numerical results}
\label{sec3}

In this section, we will present the results calculated in the QMF model
for the $\Theta^+$ in nuclear medium.
We need first to specify the parameters used in the present calculation.
For the parameters at the quark level, we take the constituent quark masses
$m_u=m_d=313\;\rm{MeV}$ and $m_s=m_{\bar{s}}=490\;\rm{MeV}$,
and the strength of the confining potential $k=700 \;\rm{MeV/fm^2}$,
as those used in the case of $\Lambda$ hypernuclei~\cite{qmf3}.
The meson masses are taken as $m_\sigma=470\;\rm{MeV}$,
$m_\omega=783\;\rm{MeV}$, and $m_\rho=770\;\rm{MeV}$.
The quark-meson couplings, $g^q_\sigma=3.14$, $g_\omega^q=4.20$
and $g_\rho^q=4.3$, and the nonlinear self-coupling constants,
$g_3=50.7$ and $c_3=53.6$, have been determined by fitting
the following equilibrium properties of nuclear matter:
equilibrium density $\rho_0 = 0.145\;\rm{fm^{-3}}$;
binding energy      $   E/A = -16.3\;\rm{MeV}    $;
incompressibility   $     k = 280  \;\rm{MeV}    $;
effective mass      $ M_N^* = 0.63 \; M_N        $;
symmetry energy     $a_{\rm{sym}}=35\;\rm{MeV}   $.
With the above parameters, the QMF model has provided quite satisfactory
results for finite nuclei and $\Lambda$ hypernuclei~\cite{qmf2,qmf3}.
Now, we take the same parameters as above to perform the calculations
for the $\Theta^+$ in nuclear medium.

It is very interesting to see how the $\Theta^+$ properties change
in nuclear matter. In Fig. 1, we present the variation of the
effective mass of the $\Theta^+$, in comparison with  those for
the nucleon and $\Lambda$, as a function of the quark mass correction
due to the presence of the $\sigma$ mean-field,
$\delta m_q=m_q-m_q^{*}=-g_{\sigma}^q\sigma$ ($q=u,d$).
It is obvious that the reduction of $M_{\Theta}^{*}$ is larger than
those of $M_N^*$ and $M_{\Lambda}^{*}$, since there are
four $u$ and $d$ quarks in the $\Theta^+$ which are influenced
by the $\sigma$ mean-field, but only three or two quarks influenced
in the nucleon or $\Lambda$. We note that the dependence of the effective
masses on the $\sigma$ mean-field must be calculated self-consistently
within the quark model, therefore the effective masses of baryons obtained
in the QMF model are not simply linear functions of $\sigma$, as given in
the relativistic mean-field (RMF) models~\cite{rmf1,rmf2}.
We show in Fig. 2 the ratios of the baryon
radii in medium to those in free space as functions of the nuclear matter
density. It is found that all of the baryons, $\Theta^+$, $\Lambda$,
and $N$ (nucleon), show significant increase of the radius in nuclear medium,
which is consistent with the famous EMC effect. The nucleon gains larger
increase of the radius than the $\Theta^+$ and $\Lambda$, because all three
quarks in $N$ are involved in the interactions with mesons
while only four-fifths in $\Theta^+$ and two-thirds in $\Lambda$ involved.
The baryon radii increase by about $4\% \sim 6\%$ at normal matter density
in the present model. In Fig. 3, we plot the scalar and vector potentials
of the $\Theta^+$ in medium as functions of the nuclear matter density,
and the results of the nucleon ($N$) and $\Lambda$ are also shown for
comparison. At $\rho=\rho_0=0.145\;\rm{fm^{-3}}$, we get
$U^{\Theta}_S=-420\;\rm{MeV}$ and $U^{\Theta}_V= 370\;\rm{MeV}$,
therefore an residual attractive  $\Theta^+$ potential,
$U^{\Theta}=U^{\Theta}_S+U^{\Theta}_V = -50\;\rm{MeV}$, is predicted
in the present model. For $\Lambda$ at $\rho=\rho_0$, we have got
the potential $U^{\Lambda}=U^{\Lambda}_S+U^{\Lambda}_V = -24\;\rm{MeV}$,
which is about half of the $\Theta^+$ potential.
In Ref.~\cite{Oset1}, a large attractive $\Theta^+$ potential
ranging from $-60$ to $-120\;\rm{MeV}$ at normal matter density
was predicted by calculating the $\Theta^+$ selfenergy
tied to the $KN$ decay of the $\Theta^+$ and the two meson baryon decay
channels of the $\Theta^+$ partners in an antidecuplet of baryons.
A QCD sum rule calculation~\cite{Tsushima} predicted an attractive
$\Theta^+$ potential of about $-40\;\sim -90\;\rm{MeV}$.
Therefore, the prediction of the $\Theta^+$ potential in the QMF model
is compatible with the results obtained by other groups~\cite{Oset1,Tsushima}.

We now present the results of self-consistent calculations
for the $\Theta^+$ hypernuclei containing a $\Theta^+$ bound in nuclei.
In Fig. 4, we show the predicted $\Theta^+$ single-particle energies
in $^{17}_{\Theta}\rm{O}$, $^{41}_{\Theta}\rm{Ca}$, and $^{209}_{\Theta}\rm{Pb}$.
It is found that there are many bound states of the $\Theta^+$ in light, medium,
and heavy nuclei. We can see that the separation between the deep $\Theta^+$
energy levels is comparable with the $\Theta^+$ width.
The theoretical prediction of the free $\Theta^+$ width in the chiral
soliton model is less than $15\;\rm{MeV}$~\cite{diakonov97},
and the $\Theta^+$ width in medium might be reduced to about one third
or less of the free width due to Pauli blocking and binding,
as pointed out in Ref.~\cite{Oset1}.
It is clear that large separation of the deep $\Theta^+$ levels
is helpful to observe distinct peaks in the experiment to produce
$\Theta^+$ bound states~\cite{Oset2}.
It is very interesting to compare the results of $\Theta^+$ hypernuclei
to those of $\Lambda$ hypernuclei. We can see that the single-particle
energy of $1s_{1/2}$ $\Theta^+$ in $^{209}_{\Theta}\rm{Pb}$ is smaller
than the one in $^{41}_{\Theta}\rm{Ca}$, which is contrary to the results
of $\Lambda$ hypernuclei. This is mainly because $\Theta^+$ carrying a positive
charge behaves like a proton in nuclei, while $\Lambda$ is a charge neutral
particle more like a neutron. The Coulomb energies of $\Theta^+$ in heavy nuclei
are much larger than those in light nuclei, that $\Theta^+$ single-particle
energies in heavy nuclei can be reduced by the large positive Coulomb energy.
We plot in Fig. 5 the scalar, vector, and Coulomb potentials
in $^{41}_{\Theta}\rm{Ca}$ with a $\Theta^+$ bound at $1s_{1/2}$ state.
At the center of the $\Theta^+$ hypernucleus,
the attractive scalar potential ($U_S^{\Theta} \sim -559\;\rm{MeV}$)
is partly cancelled by the repulsive vector potential
($U_V^{\Theta} \sim 515\; \rm{MeV}$) and Coulomb potential
($U_C \sim 11\; \rm{MeV}$), and as a consequence,
an attractive  $\Theta^+$ potential
($U^{\Theta} \sim -33\; \rm{MeV}$) remains in this case.

It is interesting and important to compare the results of the QMF model
with those obtained by other groups~\cite{Oset1,Zhong}, and discuss the
origin of the differences in the $\Theta^+$ single-particle energies.
More bound states of the $\Theta^+$ in nuclei were predicted
in some recent works~\cite{Oset1,Zhong}.
In Ref.~\cite{Oset1}, the authors obtained more deeply bound states
by solving the Schrodinger equation with two potentials:
$V(r)= -60\rho(r)/\rho_0\; (\rm{MeV})$ and
$V(r)=-120\rho(r)/\rho_0\; (\rm{MeV})$.
It is obvious that larger $\Theta^+$ single-particle energies
should be achieved due to the deeper $\Theta^+$ potentials
used in their calculations.
In Ref.~\cite{Zhong}, the $\Theta^+$ hypernuclei were studied
by using the RMF model with the $\Theta^+$ couplings,
$g^\Theta_\sigma=\frac{4}{3} g_\sigma$ and
$g^\Theta_\omega=\frac{4}{3} g_\omega$, which are the quark model
predictions close to those given in the quark-meson coupling (QMC) model.
It may be helpful to look at the case of $\Lambda$ hypernuclei,
because of the similarity of the $\Theta^+$ hypernuclei and $\Lambda$ hypernuclei.
It is well known that the properties of $\Lambda$ hypernuclei are
very sensitive to the effective coupling constants at the hadron
level, especially the two relative couplings
    $R^\Lambda_\sigma=g^\Lambda_\sigma/g_\sigma$
and $R^\Lambda_\omega=g^\Lambda_\omega/g_\omega$~\cite{rmf3}.
The quark model prediction, $R^\Lambda_\sigma=R^\Lambda_\omega=2/3$,
usually gives large overbinding of $\Lambda$ single-particle
energies in comparison with the experimental values.
For example, by taking TM1 parameter set in the RMF model with
$R^\Lambda_\sigma=R^\Lambda_\omega=2/3$, the $1s_{1/2}$ $\Lambda$
single-particle energy in $^{41}_{\Lambda}\rm{Ca}$ is calculated
to be $-35.2\;\rm{MeV}$, while in the case of the NL-SH parameter set used,
it is $-36.8\;\rm{MeV}$. We note that the experimental value of
the $1s_{1/2}$ $\Lambda$ binding energy in $^{40}_{\Lambda}\rm{Ca}$
is about $-18.7\;\rm{MeV}$.
Usually some small deviations from $R^\Lambda_\sigma=R^\Lambda_\omega=2/3$
should be taken into account in the studies of $\Lambda$ hypernuclei
in order to reproduce the experimental values~\cite{rmf3}.
It has been pointed out that about $10\%$ increase from
$R^\Lambda_\omega=2/3$ seemed to be needed to improve the overestimated
$\Lambda$ single-particle energies in the QMC model,
since the $R^\Lambda_\sigma$ calculated at the quark level
in the QMC model was very close to $2/3$~\cite{qmc1,qmc2}.
On the other hand, the QMF model yielded slightly underestimated
$\Lambda$ single-particle energies due to smaller value than $2/3$
of the $R^\Lambda_\sigma$ obtained in the QMF model, which required
about $3\%$ reduction from $R^\Lambda_\omega=2/3$ in order to
reproduce the experimental values. It should be due to the same
reason that the QMF model predicts relatively small $\Theta^+$
single-particle energies in comparison with those obtained in the
RMF model with the quark model value of the $\Theta^+$ coupling,
$g^\Theta_\sigma/g_\sigma = g^\Theta_\omega/g_\omega = 4/3$.
In the QMF model, $g^\Theta_\omega=4 g^q_\omega=\frac{4}{3} g_\omega$
is adopted according to the assumption that non-strange mesons couple
only to the $u$ and $d$ quarks inside baryons,
but $g^\Theta_\sigma=\frac{\partial M_\Theta^*}{\partial \sigma}$
has to be worked out self-consistently at the quark level.
We notice that the effective $\sigma\rm{-}baryon$ couplings
in the QMF model are not simply constants as in the RMF model.
Therefore, the relative coupling obtained in the QMF model,
$R^\Theta_\sigma=\left[\frac{\partial M_\Theta^*}{\partial \sigma}\right]
                /\left[\frac{\partial      M_N^*}{\partial \sigma}\right]$,
depends on the $\sigma$ mean-field in medium,
and we get smaller values than $4/3$ for the $R^\Theta_\sigma$
in the QMF model. This is considered as the dominant origin of
the differences between the predictions given by the QMF model
and those obtained in the RMF model~\cite{Zhong}.

In order to examine the sensitivity of $\Theta^+$ single-particle
energies to the $\Theta^+$ couplings, we perform the same
calculation with a reduced coupling $0.97\times
g_\omega^{\Theta}$. In this case, the $1s_{1/2}$ $\Theta^+$
single-particle energy in $^{41}_{\Lambda}\rm{Ca}$ changes to be
$-38.7\;\rm{MeV}$ from the value $-28.1\;\rm{MeV}$ as shown in
Fig. 4. Hence, $3\%$ reduction of $g_\omega^{\Theta}$ can lead to
$\sim 10.6\;\rm{MeV}$ decrease of the $1s_{1/2}$ $\Theta^+$
single-particle energy in $^{41}_{\Lambda}\rm{Ca}$, and it
provides $\sim 10\;\rm{MeV}$ more attraction for the $\Theta^+$
potential in nuclear matter at normal matter density. In the
present work, we have performed self-consistent calculations for
both $\Theta^+$ and nucleons in $\Theta^+$ hypernuclei. By
comparing the results of a normal nucleus containing $A$ nucleons
with those of the $\Theta^+$ hypernucleus containing $A$ nucleons
and one $\Theta^+$, the effects of the $\Theta^+$ on nucleons can
be found. The existence of the $\Theta^+$ does enlarge the scalar
and vector potentials of nucleons, but the enlargements are mostly
cancelled with each other.

\section{Conclusions}
\label{sec4}

We have studied the properties of the $\Theta^+$ in nuclear matter
and performed self-consistent calculations for $\Theta^+$ hypernuclei
in the framework of the QMF model, which has been successfully applied
to the descriptions of stable nuclei and $\Lambda$ hypernuclei.
In the present work, we have used the constituent quark model
to describe the $\Theta^+$ and nucleons, which naturally allows
the direct coupling of exchanged mesons with quarks inside baryons.
With the parameters determined by the equilibrium properties of nuclear
matter and the assumption that non-strange mesons couple only to
the $u$ and $d$ quarks, an attractive $\Theta^+$ potential has been
achieved as a consequence of the cancellation between the attractive
scalar potential and the repulsive vector potential in nuclear matter.
The QMF model provides an attractive $\Theta^+$ potential of
$\sim -50\;\rm{MeV}$, and yields a reduction of effective $\Theta^+$
mass of $\sim 420\;\rm{MeV}$ as well as $\sim 5\%$ increase of the
$\Theta^+$ radius at normal matter density.

We have investigated the $\Theta^+$ single-particle energies in light,
medium, and heavy nuclei. It is found that there are many bound states
in $\Theta^+$ hypernuclei, and the $\Theta^+$ seems to be more deeply
bound than the $\Lambda$ in hypernuclei. We have discussed the
sensitivity of $\Theta^+$ single-particle energies to the $\Theta^+$
couplings. The comparison between the results calculated in the QMF model
and those obtained by other groups has been done, and the origin of the
differences has also been discussed by referring to the case of
$\Lambda$ hypernuclei. The results obtained in the QMF model provide
encouragement for the experimental work to produce $\Theta^+$ hypernuclei.

We comment here a possible method to produce $\Theta^+$
hypernuclei.  The gamma induced reaction on nucleus for the
formation of the $\Theta^+$ leads to the $\Theta^+$ in a
scattering state due to the large momentum transfer.  On the other
hand, the $K^+$ induced reaction with a nucleus could produce the
$\Theta^+$ with a reasonable momentum of order $500\;\rm{MeV}$, when an
outgoing pion is observed at the forward angle.  This momentum
transfer would lead to high angular momentum transfer of order
$L \simeq 10$, when the $\Theta^+$ is produced at the nuclear surface
with a medium heavy nuclei.  Hence, it is highly plausible to
observe a sharp peak with high angular momentum; high spin
$\Theta^+$ - high spin nucleon hole state, popping up in the
beginning of the continuum region as the case of the (p,$\pi^-$)
reaction~\cite{brown}.

\section*{Acknowledgments}

This work was supported in part by the National Natural Science
Foundation of China (No. 10135030) and the Specialized Research Fund
for the Doctoral Program of Higher Education (No. 20040055010).



\begin{figure}[htb]
\vspace{0.1cm}
\epsfxsize=16cm
\epsfysize=24cm
\centerline{\epsfbox{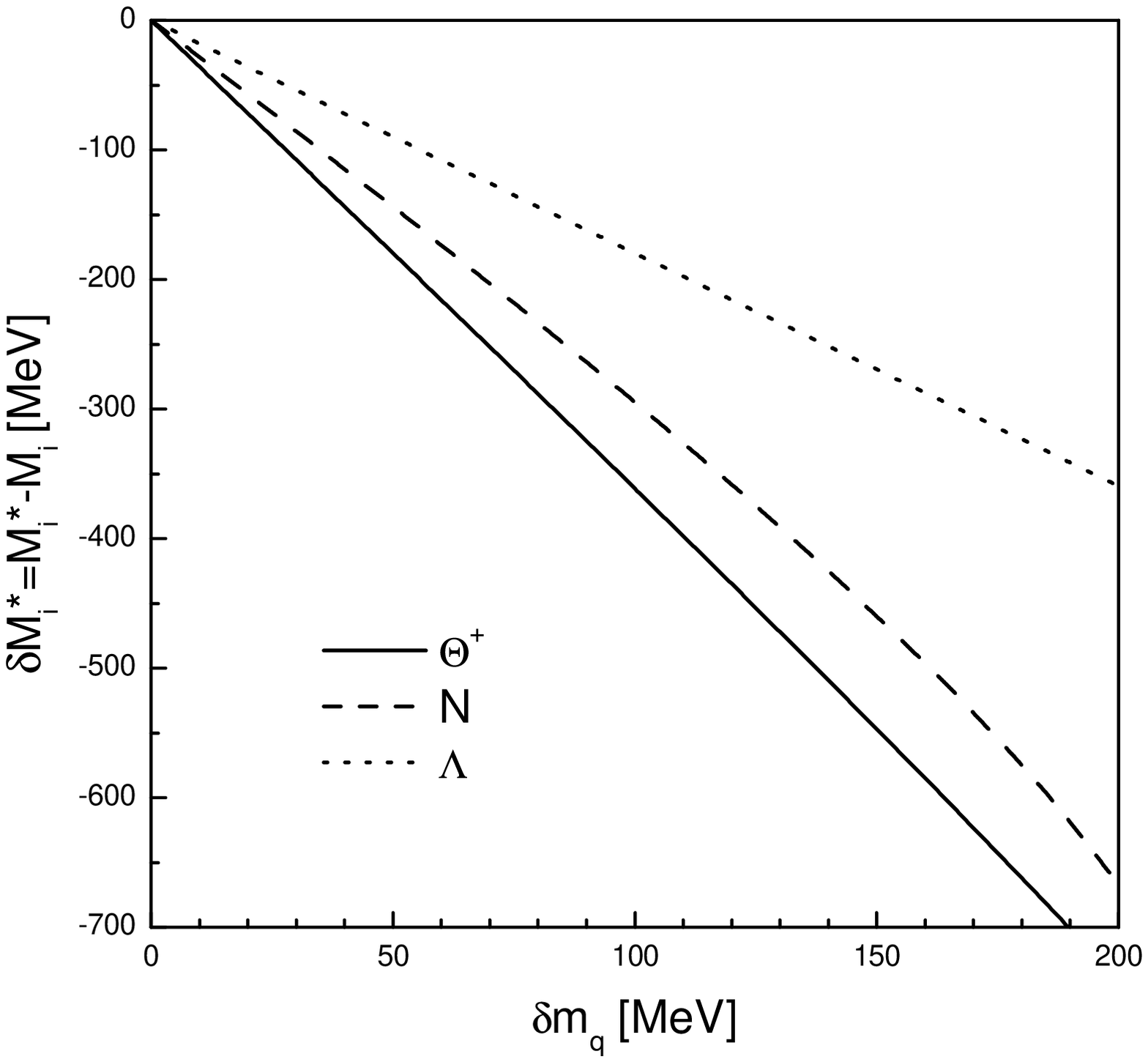}}
\vspace{-6.0cm}
\caption{The variations of the baryon effective masses,
         $\delta M_i^*=M^*_i-M_i$ ($i=N,\Lambda,\Theta^+$),
         as functions of the quark mass correction,
         $\delta m_q=m_q-m_q^{*}=-g_{\sigma}^q\sigma$.}
\end{figure}

\begin{figure}[htb]
\vspace{0.1cm}
\epsfxsize=16cm
\epsfysize=24cm
\centerline{\epsfbox{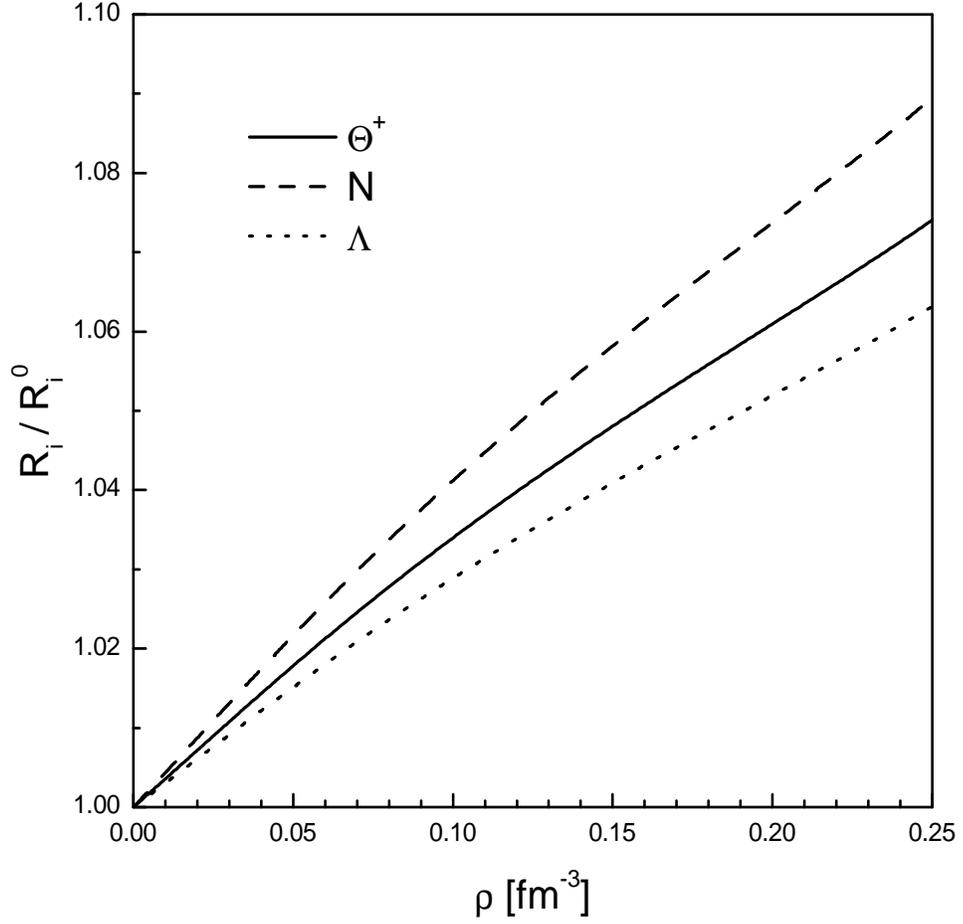}}
\vspace{-6.0cm}
\caption{The ratios of the baryon radii in nuclear matter $R_i$
    to that in free space $R_i^0$ as functions of the nuclear
    matter density $\rho$.}
\end{figure}

\begin{figure}[htb]
\vspace{0.1cm}
\epsfxsize=16cm
\epsfysize=24cm
\centerline{\epsfbox{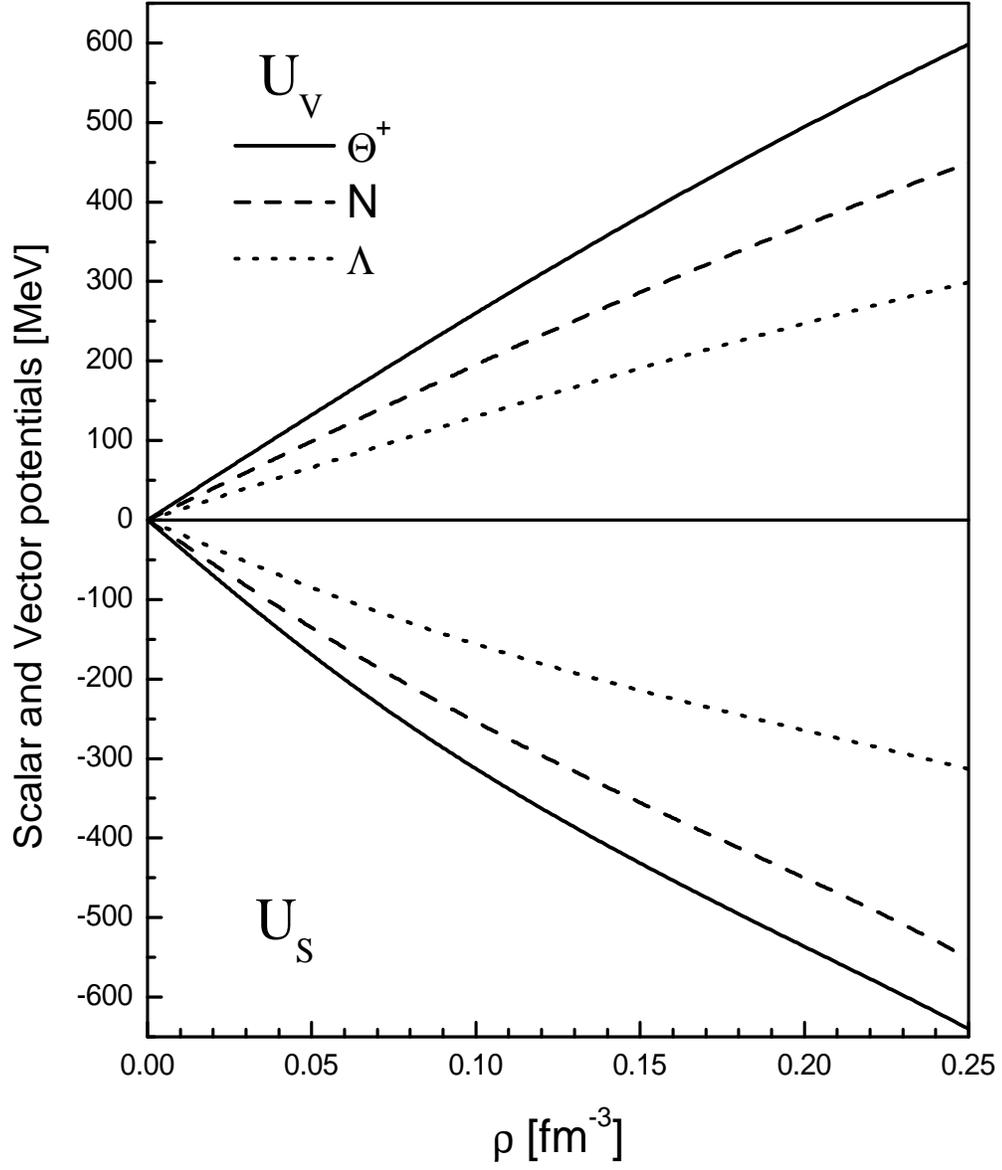}}
\vspace{-5.0cm}
\caption{The scalar and vector potentials, $U_S$ and $U_V$,
         as functions of the nuclear matter density $\rho$.}
\end{figure}

\begin{figure}[htb]
\vspace{0.1cm}
\epsfxsize=16cm
\epsfysize=24cm
\centerline{\epsfbox{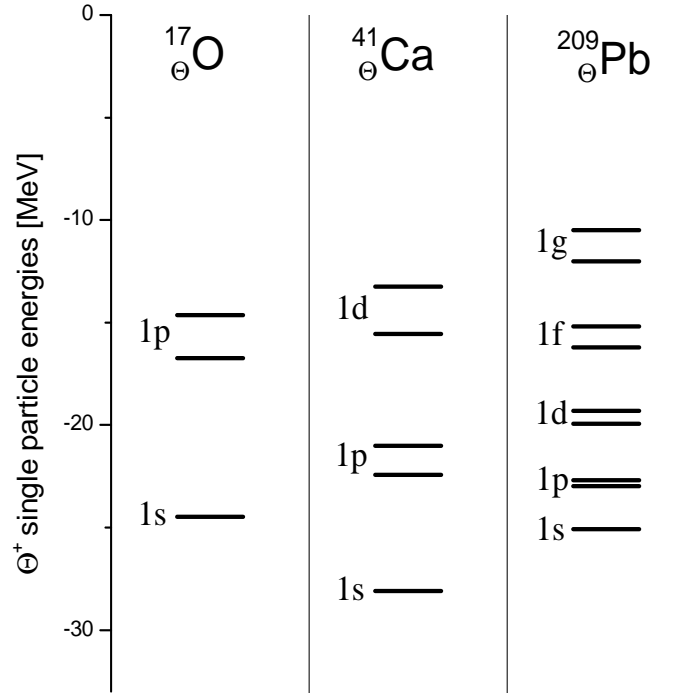}}
\vspace{-6.0cm}
\caption{The $\Theta^+$ single-particle energies in
         $^{17}_{\Theta}\rm{O}$, $^{41}_{\Theta}\rm{Ca}$,
         and $^{209}_{\Theta}\rm{Pb}$.}
\end{figure}

\begin{figure}[htb]
\vspace{0.1cm}
\epsfxsize=16cm
\epsfysize=24cm
\centerline{\epsfbox{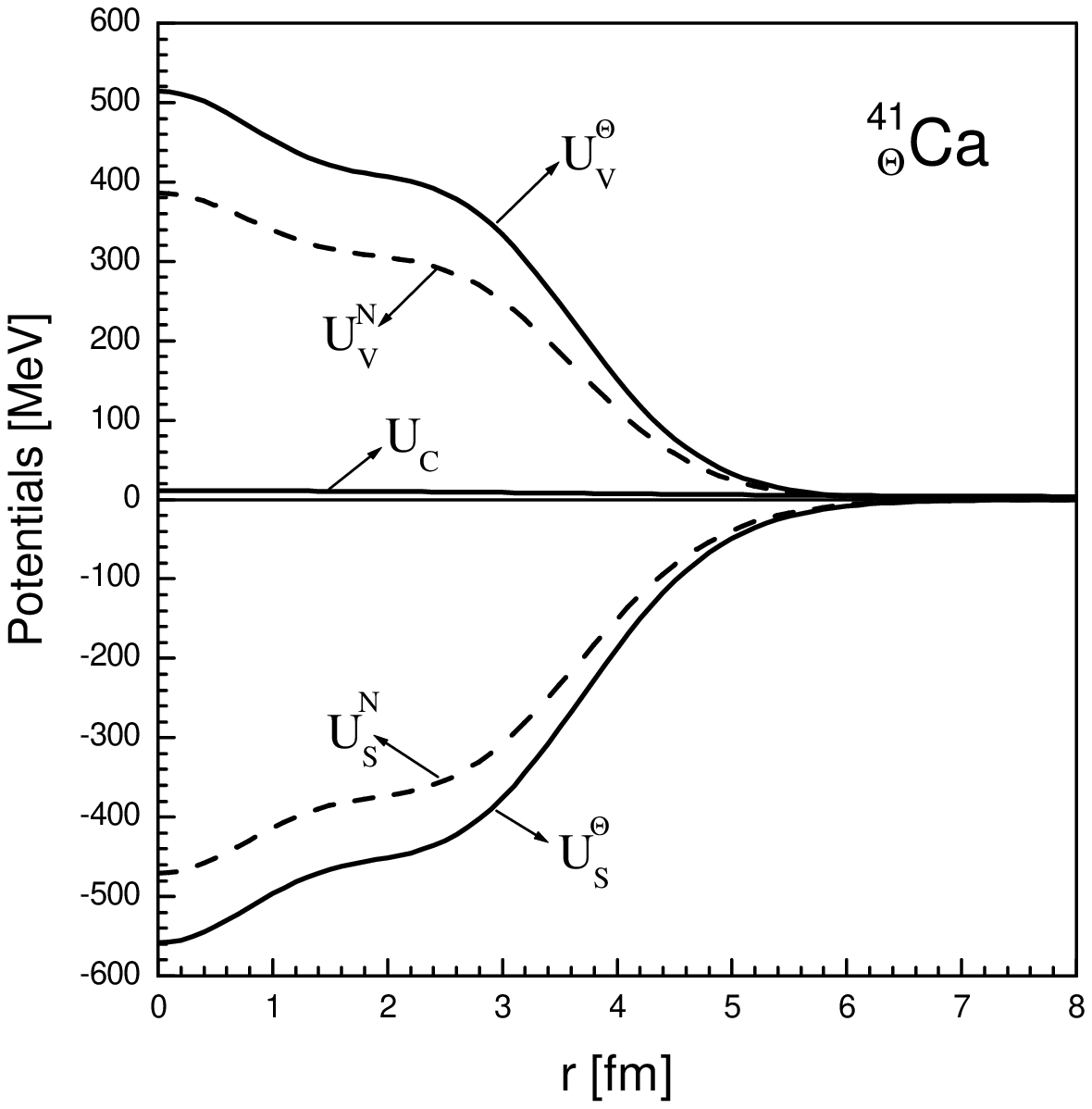}}
\vspace{-6.0cm}
\caption{The scalar potential $U^i_S$, the vector potential $U^i_V$,
         and the Coulomb potential $U_C$ for the nucleon ($i=N$)
         and the $\Theta^+$ ($i=\Theta$)
         in $^{41}_{\Theta}\rm{Ca}$ with a $\Theta^+$ bound at $1s_{1/2}$ state.}
\end{figure}


\newpage
\begin{thebibliography}{99}

\bibitem{nakano02} T. Nakano et al. (LEPS collaboration), Nucl. Phys. {\bf A721}, 112c (2003).

\bibitem{nakano03} T. Nakano et al. (LEPS collaboration), Phys. Rev. Lett. {\bf 91}, 012002 (2003).

\bibitem{diakonov97} D. Diakonov, V. Petrov, and M. Polyakov, Z. Phys. A {\bf 359}, 305 (1997).

\bibitem{penta04} http://www.rcnp.osaka-u.ac.jp$/^\sim$penta04/

\bibitem{Miller} G.A. Miller, Phys. Rev. C {\bf 70}, 022202 (2004).

\bibitem{Kim} H.C. Kim, C.H. Lee, and H.J. Lee, J. Kor. Phys. Soc. {\bf 46}, 393 (2005).

\bibitem{Oset1} D. Cabrera, Q.B. Li, W. Magas, E. Oset, and M.J. Vicente Vacas, Phys. Lett. B {\bf 608}, 231 (2005).

\bibitem{Tsushima} F.S. Navarra, M. Nielsen,  and K. Tsushima, Phys. Lett. B {\bf 606}, 335 (2005).

\bibitem{Zhong} X.H. Zhong, Y.H. Tan, L. Li, and P.Z. Ning, Phys. Rev. C {\bf 71}, 015206 (2005).

\bibitem{Oset2} H. Nagahiro, S. Hirenzaki, E. Oset, and M.J. Vicente Vacas, nucl-th/0408002.

\bibitem{qmf2} H. Shen and H. Toki, Phys. Rev. C {\bf 61}, 045205 (2000).

\bibitem{qmf3} H. Shen and H. Toki, Nucl. Phys. {\bf A707}, 469 (2002).

\bibitem{rmf1} B.D. Serot and J.D. Walecka, Adv. Nucl. Phys. {\bf 16}, 1 (1986).

\bibitem{rmf2} Y. Sugahara and H. Toki, Prog. Theor. Phys. {\bf 92}, 803 (1994).

\bibitem{rmf3} C.M. Keil, F. Hofmann, and H. Lenske, Phys. Rev. C {\bf 61}, 064309 (2000).

\bibitem{qmc1} K. Tsushima, K. Saito, and A.W. Thomas, Phys. Lett. B {\bf 411}, 9 (1997);
               (E) Phys. Lett. B {\bf 421}, 413 (1998).

\bibitem{qmc2} K. Tsushima, K. Saito, J. Haidenbauer, and A.W. Thomas,
               Nucl. Phys. {\bf A630}, 691 (1998).

\bibitem{brown} A.B. Brown, O. Scholten, and H. Toki, Phys. Rev. Lett. {\bf 51}, 1952 (1983).

\end{thebibliography}
\end{document}